\documentclass[reprint,superscriptaddress,amsmath,amssymb,aps,prb,longbibliography,
]{revtex4-2}

\usepackage[dvipsnames]{xcolor}
\definecolor{mblue}{RGB}{31, 119, 180}
\usepackage{hyperref}
\hypersetup{backref,
colorlinks=true,breaklinks,urlcolor=mblue,linkcolor=mblue,citecolor=mblue}
\usepackage{physics, braket, bm, amsthm, amsmath, amssymb, mathrsfs, graphicx, dcolumn, subfigure}
\usepackage{mathrsfs}
\usepackage[]{ulem}

\begin{document}
\title{Machine-learning-inspired quantum control in many-body dynamics}
\author{Meng-Yun Mao}
\affiliation{College of Physics, Nanjing University of Aeronautics and Astronautics, Nanjing 211106, China}
\affiliation{Key Laboratory of Aerospace Information Materials and Physics (NUAA), MIIT, Nanjing 211106, China}

\author{Zheng Cheng}
\affiliation{College of Physics, Nanjing University of Aeronautics and Astronautics, Nanjing 211106, China}
\affiliation{Key Laboratory of Aerospace Information Materials and Physics (NUAA), MIIT, Nanjing 211106, China}

\author{Liangsheng Li}
\affiliation{National Key Laboratory of Scattering and Radiation, Beijing 100854, China}

\author{Ning Wu}
\email{wunwyz@gmail.com}
\affiliation{Center for Quantum Technology Research, School of Physics, Beijing Institute of Technology, Beijing 100081, China}
\affiliation{Key Laboratory of Advanced Optoelectronic Quantum Architecture and Measurements (MOE), School of Physics, Beijing Institute of Technology, Beijing 100081, China}

\author{Wen-Long You$\,$}
\email{wlyou@nuaa.edu.cn}
\affiliation{College of Physics, Nanjing University of Aeronautics and Astronautics, Nanjing 211106, China}
\affiliation{Key Laboratory of Aerospace Information Materials and Physics (NUAA), MIIT, Nanjing 211106, China}

\date{\today}
\begin{abstract}
Achieving precise preparation of quantum many-body states is crucial for the practical implementation of quantum computation and quantum simulation. 
However, the inherent challenges posed by unavoidable excitations at critical points during quench processes necessitate careful design of control fields. 
In this work, we introduce a promising and versatile dynamic control neural network tailored to optimize control fields. 
We address the problem of suppressing defect density and enhancing cat-state fidelity during the passage across the critical point in the quantum Ising model.  
Our method facilitates seamless transitions between different objective functions by adjusting the {optimization strategy}. In comparison to gradient-based power-law quench methods, our approach demonstrates significant advantages for both small system sizes and long-term evolutions.
We provide a detailed analysis of the specific forms of control fields and summarize common features for experimental implementation. 
Furthermore, numerical simulations demonstrate the robustness of our proposal against random noise and spin number fluctuations. The optimized defect density and cat-state fidelity exhibit a transition at a critical ratio of the quench duration to the system size, coinciding with the quantum speed limit for quantum evolution.

\end{abstract}

\maketitle

\section{Introduction}

Nowadays it is imperative to achieve precise preparation of quantum many-body states due to the urgent demand for applications in various fields, including quantum computation~\cite{farhi_quantum_2001, PhysRevLett.102.220501} and quantum simulation~\cite{PhysRevLett.93.207204,PhysRevA.81.061603,PRXQuantum.3.030343}. 
In particular, quantum metrology, a rapidly advancing field of quantum information science, is currently experiencing a surge of theoretical developments~\cite{PhysRevLett.124.110502, PhysRevLett.121.020402, PhysRevLett.127.060501} and experimental breakthroughs~\cite{PhysRevLett.112.150802, hudelist_quantum_2014, marciniak_optimal_2022, yin_experimental_2023}. A pivotal focus of a quantum metrology scheme involves the preparation of optimal nonclassical states~\cite{Toth_2014,szigeti2021improving}.
Through the utilization of inherent quantum properties, such as entanglement
~\cite{science.1104149, giovannetti_advances_2011, huang2014quantum}, coherence~\cite{PhysRevLett.123.180504} and discord~\cite{PhysRevLett.105.020503, PhysRevLett.88.017901}, 
quantum metrology can achieve the sensitivity for estimating unknown parameters that surpasses the constraints imposed by the standard quantum limit   of classical strategies \footnote{If there are $N$ qubits in separable states, it is known that 
the uncertainty
(that is, the inverse of the sensitivity) scales as ${\cal O}(N^{-1/2})$, which
is called the standard quantum limit, while quantum physics allows the ultimate scalings being ${\cal O}(N^{-1})$, i.e., the Heisenberg scaling, in the absence of noise and ${\cal O}(N^{-3/4})$ in the presence of realistic decoherence}. 
In fact, the enhanced sensitivity may even approach the  limits set by the Heisenberg uncertainty principle~\cite{PhysRevLett.128.200501}. To this end, various forms of entangled states have been
generated in engineered many-body systems to increase the phase sensitivity~\cite{lucke_twin_2011, strobel_fisher_2014}.  However, a comprehensive understanding of the relationship between quantum features and ultimate scaling sensitivity beyond the standard quantum limit remains elusive. 

A widely employed method for the theoretical~\cite{PhysRevLett.107.016402, PhysRevLett.126.103401, PhysRevB.91.134303, PhysRevLett.117.040501, PhysRevResearch.5.L022037} and experimental~\cite{motta_determining_2020, choi_preparing_2023, leonard_realization_2023, bernien_probing_2017, maslova_probing_2019} preparation of quantum many-body states involves utilizing unitary evolution. This process transforms the initial state, typically the ground state of some simple Hamiltonian $\hat{H}_i$, into the ground state of the target Hamiltonian $\hat{H}_f$. 
According to the adiabatic theorem,
the unitary transformation can be implemented with arbitrary accuracy by changing the Hamiltonian sufficiently slowly~\cite{JPSJ, PhysRevApplied.15.044043}.
Nevertheless, the strict requirement of the adiabatic limit dictates infinitely long evolution time,  inevitably resulting in system decoherence and the disappearance of entanglement~\cite{PhysRevLett.87.097901}. In particular, 
crossing a quantum critical point (QCP) in finite time challenges the adiabatic condition due to the closing of the
energy gap, which ultimately results in the formation of excitations~\cite{Jacek_2010}.
Consequently, a trade-off must be considered between the evolution time and the excitations generated by the system. In this context, tremendous efforts have recently been dedicated to achieving an optimal passage through the non-adiabatic evolution, particularly across the QCP,
aiming to minimize unwanted excitations~\cite{PhysRevLett.101.076801, PhysRevLett.95.245701} or to maximize fidelity~\cite{jozsa_fidelity_1994, nielsen2002quantum}.
Shortcut to adiabaticity (STA) provides a way of finding fast trajectories that connect the initial and final states by manipulating the system's parameters in a non-adiabatic fashion while still obtaining results akin to those of an adiabatic process~\cite{PhysRevLett.109.115703, PhysRevB.90.104306, Damski_2014, PhysRevX.4.021013, PhysRevLett.114.177206}. Given the flexibility in selecting intermediate trajectories, the time-dependent control parameters of a system can be adjusted in various STA protocols~\cite{RevModPhys.91.045001}.
One common method of STA involves introducing counterdiabatic driving to the reference Hamiltonian that effectively compensates for the non-adiabatic behavior.
However, implementing this technique in quantum many-body systems remains a challenging task for experimental execution.

Alternatively, a prevalent approach is quantum optimal control~\cite{koch_quantum_2022, PhysRevLett.106.190501, PhysRevLett.111.260501, PhysRevLett.103.240501, GIANNELLI2022128054}, where time-dependent control parameters of a system are fine-tuned using optimal control theory. 
Governed by time-energy uncertainty relations, the characteristic time scales during non-adiabatic evolution are encapsulated by the quantum speed limit~\cite{PhysRevA.67.052109, PhysRevLett.103.160502, Deffner_2017}, 
which delves into the minimum time required for quantum states to achieve specific predetermined objectives. 
This is particularly crucial in the field of quantum information, where rapid dynamics are often advantageous~\cite{PhysRevX.9.011034, zhang_quantum_2023, Deffner_2017}.
While the quantum optimal control finds widespread applications in various systems, it is of fundamental interest to formulate the control theory in a general framework. Over the
past few years, machine-learning technology has become an integral part of the optimization theory and has been proved to be applicable to optimizing the parameters of variational states in a variety of interacting quantum many-body systems
~\cite{csp_355, PhysRevLett.121.167204, choo_fermionic_2020, zhu_flexible_2022, carleo_constructing_2018, PhysRevLett.122.020601}. 
In this work, we propose a promising and generalizable dynamic control neural network (DCNN) to optimize the control field passing through two fixed points of the quantum critical systems that holds potential for testing and comparing approximate methods, as well as bearing implications for condensed matter physics. To demonstrate the power of our method, we focus on suppressing the defect density and improving the cat-state fidelity during the passage across the critical point in the quantum Ising model. 
Compared to the gradient-based power-law quench~\cite{PhysRevB.91.041115}, the method offers significant advantages over small system size and long-term evolution. We also numerically demonstrate that our proposal is robust against the random noise and the spin number fluctuations. The optimized defect density and cat-state fidelity are uncovered to coincide with the quantum speed limit for quantum evolution.

The rest of the paper is organized as follows.
In Sec.~\ref{method} we 
present the gradients of specific observables with respect to the time-dependent control parameter in the many-body systems that can be transformed into free fermions. 
Additionally, we introduce  the DCNN method in this section. 
Section~\ref{optimal} is devoted to the application of the DCNN to the suppression of the defect density in the quantum Ising model.  A thorough analysis of achieving the {optimized} control fields is also discussed. The effectiveness of the protocol, particularly under varying conditions such as random noise and fluctuations in the number of spins, is evaluated through comprehensive numerical simulations. 
Section \ref{sec_catstate}  delves into the improvement of the cat-state fidelity in the quantum Ising model.
Finally, the conclusion and outlook are given in Sec.~\ref{conclusion}.

\section{Dynamical protocol in integrable many-body models} \label{method}
\subsection{Quantum Ising-like models} 
The quantum Ising model has emerged as a prototypical model of quantum many-body systems due to its analytical solvability.   
It thus serves as a valuable tool for exploring emerging quantum phenomena~\cite{mi_time-crystalline_2022, abm7652, PhysRevResearch.4.043027, abq5769, PhysRevB.98.155134, PhysRevLett.127.100504, PhysRevLett.125.260505}
and as a benchmark for evaluating the effectiveness of state-of-the-art approaches and algorithms~\cite{PRXQuantum.1.020320, PRXQuantum.3.020347, PhysRevLett.127.020502}.
Furthermore, it is an especially compelling option as a versatile platform for quantum simulation~\cite{PhysRevLett.107.250503, Schauss_2018, kim_quantum_2010, PhysRevLett.101.220501}. 
The exact simulation of Ising model has been implemented on quantum computers~\cite{CerveraLierta2018exactisingmodel,kim_evidence_2023}.  Actually, the steps to build quantum circuits at a scale that could provide an advantage for simulating the Ising Hamiltonian follow the same strategy as the analytical solution of the models. Therefore, the method can
be extended to other integrable models. 
To this end, we focus on a class of integrable many-body spin systems that can be mapped into free fermion models. 
A variety of systems holding quantum phase transitions can be mapped into such {kinds} of models,
including the one-dimensional (1D) quantum Ising model and XY model,
which is regarded as one of two canonical quantum critical systems~\cite{Sachdev_1999},
as well as the 1D and two-dimensional Kitaev models~\cite{KITAEV20032, WU20123530}, etc.
These cover a variety of spin models that are of interest in both quantum information science and condensed matter physics. 

\par Consider a closed quantum system which depends on time explicitly through a control field $g(t)$ that can be varied arbitrarily with certain constraints. The control field $g(t)$ is assumed to enter $\hat{H}[g(t)]$ via a term $g(t) \hat{Y}$ with $\hat{Y}$ some time-independent operator.
We consider a quench starting from some fixed initial value $g_{i} = g(-T)$ and ending up with a fixed final value $g_{f} = g(T)$ within the time interval $t \in [-T , T]$. In this work, we mainly focus on quantum systems that themselves belong to, or can be mapped to, $d$-dimensional free-fermion models. 
In general, this family of arbitrary-dimensional time-dependent free-fermion Hamiltonians can be written as
\begin{equation}
    \hat{H}(t)=\sum_{\vec{k}} \psi_{\vec{k}}^{\dagger} [\vec{d}_{\vec{k}}(g(t)) \cdot \vec{\sigma}_{\vec{k}}] \psi_{\vec{k}},
    \label{eq:H}
\end{equation}
where $\vec{\sigma}_{\vec{k}} = (\sigma_{\vec{k}}^{x} , \sigma_{\vec{k}}^{y} , \sigma_{\vec{k}}^{z})$
are the Pauli matrices acting on the mode $\vec{k}$ and $\psi_{\vec{k}} = (a_{\vec{k}},b_{\vec{k}})^{T}$ are certain fermionic operators.
The function $\vec{d}_{\vec{k}}(g) = (d_{\vec{k}}^{x}(g) , d_{\vec{k}}^{y}(g) , d_{\vec{k}}^{z}(g))$ is determined by specific models and the corresponding norm $\varepsilon_{\vec{k}} = |\vec{d}_{\vec{k}}(g)|= \sqrt{d_{\vec{k}}^{x}(g)^{2} + d_{\vec{k}}^{y}(g)^{2} + d_{\vec{k}}^{z}(g)^{2}}$ gives the single-particle dispersion.
Equation~\eqref{eq:H} can be diagonalized directly within each $\vec{k}$-subspace as $\hat{H}(t) = \sum_{\vec{k}} \hat{H}_{\vec{k}}(t)$ with
\begin{equation}
    \hat{H}_{\vec{k}}(t) = \varepsilon_{\vec{k}}(\Psi_{\vec{k}}^{\dagger} \Psi_{\vec{k}} - 1),
\end{equation}
where $\Psi_{\vec{k}} = (A_{\vec{k}},B_{\vec{k}})^{T}$ is associated with $\psi_{\vec{k}}$ by certain $\vec{k}$ dependent unitary transformation.
Note that both the dispersion $\varepsilon_{\vec{k}}$ and the quasi-fermion operators $\Psi_{\vec{k}}$ are generally time-dependent through the control field $g(t)$.
The sub-ground state $| G_{\vec{k}}[g(t)] \rangle$ of mode $\vec{k}$ satisfies
\begin{equation}
    A_{\vec{k}} | G_{\vec{k}}[g(t)] \rangle = B_{\vec{k}} | G_{\vec{k}}[g(t)] \rangle = 0.
\end{equation}
The sub-excited state with the same fermion number parity as $G_{\vec{k}}[g(t)]$ is obviously
\begin{equation}
    | \bar{G}_{\vec{k}}[g(t)] \rangle = A_{\vec{k}}^{\dagger}(t) B_{\vec{k}}^{\dagger}(t) | G_{\vec{k}}[g(t)] \rangle.
\end{equation}
We assume that the ground state is always nondegenerate in the process of the quench and the energy gap is closed only at the QCP in the thermodynamic limit. Here, both the global instantaneous ground state $|G[g(t)]\rangle$ and time evolution operator $U(t , -T)$ are separable,
\begin{eqnarray}
    | G[g(t)] \rangle & = & \prod_{\vec{k}} | G_{\vec{k}}[g(t)] \rangle , \\
    U(t , -T) & = & \prod_{\vec{k}} U_{\vec{k}}(t , -T),
\end{eqnarray}
where $U_{\vec{k}}(t , -T) = \mathcal{T} \exp (i \int_{-T}^{t} d\tau H_{\vec{k}}(\tau))$ with $\mathcal{T}$ the time-ordered operator. The time-evolved state is given by $| \phi(t) \rangle = U(t , -T) | G(g_{i}) \rangle$, which is also separable, i.e., $| \phi(t) \rangle=\prod_{\vec{k}}U_{\vec{k}}(t,-T)|G_{\vec{k}}(g_i)\rangle$.

\subsection{Controlling observables and their gradients. } Our aim is to minimize or maximize the final expectation value $O(T)\equiv \langle \phi(T)| \hat{O}  | \phi(T) \rangle$ of a general observable $\hat{O}$.  As will be discussed below, a key quantity utilized in the DCNN is the gradient of $O(T)$ with respect to the control field $g(t)$ \cite{Brif_2010} (see appendix~\ref{AppA}),
\begin{eqnarray}\label{dOdg1}
\frac{\delta O(T)}{\delta g(t)}=2\Im\langle G(g_i)|\hat{O}(T)\hat{Y}(t)|G(g_i)\rangle,
\end{eqnarray}
where $\hat{A}(t)=U^\dag(t,-T)\hat{A}U(t,-T)$ is the Heisenberg-picture operator for any operator $\hat{A}$.

\par We further assume that 
the control term $\hat{Y}$ can be expressed as a summation over even operators of independent modes
\begin{equation}
    \hat{Y} = \sum_{\vec{k}} \hat{Y}_{\vec{k}}.
\end{equation}
Here we are interested in the following two types of observables, i.e., $\hat{O}_1=\sum_{\vec{k}}\hat{O}_{1,\vec{k}}$ and $\hat{O}_2=\prod_{\vec{k}}\hat{O}_{2,\vec{k}}$.
The gradient of $O_1(T)$ has been calculated as (see Appendix~\ref{AppA} for detail) 
\begin{eqnarray}\label{grad1}
\frac{\delta O_1(T)}{\delta g(t)}=2\Im\sum_{\vec{k}}\langle\phi_{\vec{k}}(T)|\hat{O}_{1,\vec{k}}|\bar{\phi}_{\vec{k}}(T)\rangle\langle\bar{\phi}_{\vec{k}}(t)|\hat{Y}_{\vec{k}}|\phi_{\vec{k}}(t)\rangle,\nonumber\\
\end{eqnarray}
where
$|\bar{\phi}_{\vec{k}}(t)\rangle=U_{\vec{k}}(t,-T)|\bar{G}_{\vec{k}}(g_i)\rangle$
is the evolved sub-excited state of mode $\vec{k}$. The gradient of $O_2(T)$ can be calculated in a similar way, 
\begin{eqnarray}\label{grad2}
\frac{\delta O_2(T)}{\delta g(t)}&=&2\prod_{\vec{p}}\langle\phi_{\vec{p}}(T)|\hat{O}_{2,\vec{p}}|\phi_{\vec{p}}(T)\rangle\times\nonumber\\
&&\sum_{\vec{k}}\frac{\Im\langle\phi_{\vec{k}}(T)|\hat{O}_{2,\vec{k}}|\bar{\phi}_{\vec{k}}(T)\rangle\langle\bar{\phi}_{\vec{k}}(t)|\hat{Y}_{\vec{k}}|\phi_{\vec{k}}(t)\rangle}{\langle\phi_{\vec{k}}(T)|\hat{O}_{2,\vec{k}}|\phi_{\vec{k}}(T)\rangle}.\quad\quad
\end{eqnarray}

\subsection{ Dynamic control neural network}  
We then introduce the DCNN method
to {optimize the observables in many-body dynamics}.
As an extension of the control neural network~\footnote{Zheng Cheng, Meng-Jiao Lyu, Takayuki Myo, Hisashi Horiuchi, Hiroshi Toki, Zhong-Zhou Ren, Masahiro Isaka, Meng-Yun Mao, Hiroki Takemoto, Niu Wan, Wen-Long You, and Qing Zhao, to appear}, the DCNN is a neural network with an adaptive dynamic structure and the flexible {optimization strategy}. 
  This methodology not only allows us to design more diverse control fields but also facilitates the further minimization of specific observables, a capability not fully realized by other methods, 
e.g., the direct gradient algorithm {based on the power-law quench}~\cite{PhysRevB.91.041115}.
To be concrete, the DCNN possesses a similar architecture to the conventional neural network,
which is composed of input, hidden, and output layers.
Two adjacent layers are connected by the weights, biases, and activation function. The distinguishing features 
of the DCNN lie in its dynamic structure of the network and the learning {optimization strategy}.
The dynamic structure of the network can be 
effectively executed by increasing the number of units in each hidden layer. 
We opt for two hidden layers and $\tanh(x)$ as the activation function. 
\begin{figure*}
    \includegraphics[width=0.9\linewidth]{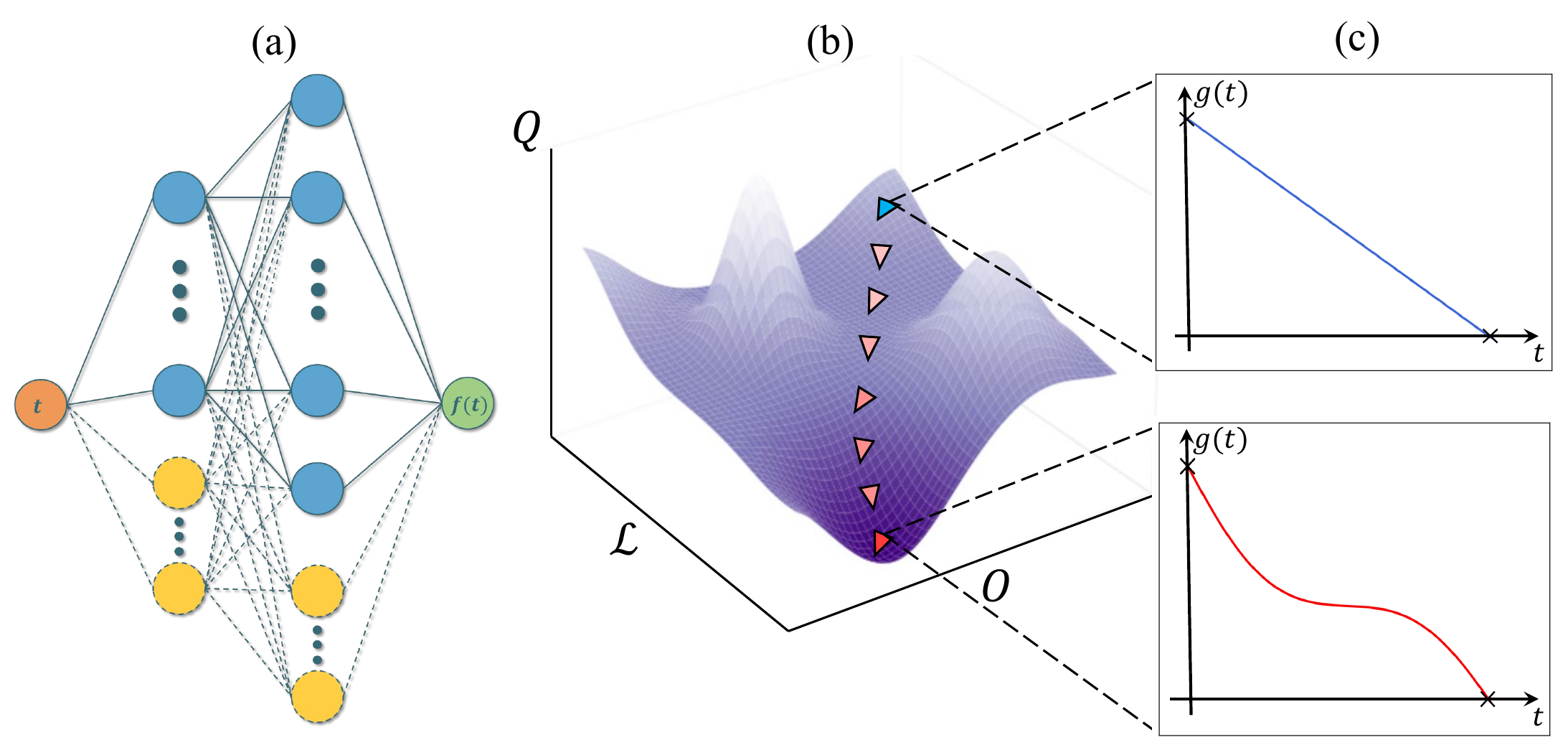}
    \caption{The workflow of the DCNN and the visualization of the optimization landscape. 
    (a) The structure of the DCNN. We choose two hidden layers with $n_{1}$ units in the first layer and $n_{2}$ in the second.
    $t$ is the input of the neural network and $f(t)$ is the output.
    Each neuron in the hidden and output layers has a corresponding weight and bias. 
    The blue dots are the original units in the neural network, the yellow dots are the newly increased units
    and the dashed lines represent the corresponding increased connection.
    (b) The optimization of the combined objective function $Q$ for the observable  $O$ and the loss $\mathcal{L}$. An initial trial form of the control field is used as a starting point (blue triangle).  
    $Q$ moves downhill  (darker red triangles) until the convergence is reached.
    To be noticed, the downhill movement signifies a simultaneous decline in both $O$ and $\mathcal{L}$.
    (c) The initial and final forms of the control field $g(t)$ with two fixed points (black crosses). 
    }
    \label{fig_DCNN}
\end{figure*}

The workflow of the DCNN is illustrated in Fig.~\ref{fig_DCNN}.
The neural network has two hidden layers with $n_{1}$ units in the first layer and $n_{2}$ in the second. The parameters of the neural network can be randomly initialized.
For simplicity, the parameters can be initialized by training the neural network based on the empirical form of the control field. We then add {$\delta n_{1}$} neurons to the first hidden layers and add {$\delta n_{2}$} to the second.
The corresponding parameters of the newly added neurons should be randomly initialized based on the average distribution.
The input of the DCNN is $t$ and the output is the following specific function
\begin{eqnarray}
    f(t) = W_3^{\rm{T}} \tanh. [W_2 \tanh. (W_1 t + B_1) + B_2],
    \label{eq:ft}
\end{eqnarray}
where $W_{i} \ (i=1 , 2 , 3)$ are weight matrices,
$B_{i} \ (i=1 , 2)$ are bias matrices,
and $\tanh.(x)$ is the function acting on each elements of the matrix.
{The output $f(t)$ undergoes a linear transformation to produce the control fields $g(t)$. Interestingly, increasing the number of neurons within the neural-network-based approach exhibits a conceptual parallel in model space expansion, akin to the methods employed in traditional optimal control that augment the number of optimization parameters~\cite{PhysRevA.92.062343, PhysRevA.94.023420}.
Nevertheless, the DCNN adaptively and flexibly explores model space, unlike the traditional optimal control method that enhances it by requiring prior knowledge and specific basis functions.}
{
The DCNN employs a pareto-optimization-based strategy, adeptly balancing multiple objectives without compromising any particular target. 
}
Due to the flexible {optimization strategy} of the DCNN, we can impose desired constraints on the form of the control field. 
In our work, 
the control field $g(t)$ has fixed values for initial moment $t=-T$ and final moment $t=T$, 
which can be evaluated using the loss function
\begin{eqnarray}
    \mathcal{L} = \Big( g(-T) - r \Big)^{2} + \Big( g(T) - s \Big)^{2},
\end{eqnarray}
where $r$ and $s$ are the fixed points of $g(t)$ for $t=-T$ and $T$, respectively.
The other constraint, involving the extremum of the final expectation value of a specific observable $O(T)$.

{The DCNN is capable of conducting the optimization for multiple objectives, aligning with approaches in quantum  optimal control~\cite{PhysRevA.78.033414}. }
To be specific, the classical approach to solve a multi-objective optimization problem is to assign a weight $q_{i}$ to each normalized objective function so that the problem is converted to a single objective problem with a scalar objective function $Q$~\cite{GIAGKIOZIS2015338}. 
In our approach, the objective function is not necessarily normalized.
We set $Q$ to be $Q = q_{1} O  + q_{2} \mathcal{L}$,
where $q_{1}$ and $q_{2}$ are two arbitrary real numbers. {Notably, the normalization of the individual objective functions is not imperative, given the adaptability provided by the adjustable coefficients $q_i$ ($ i$ = 1 , 2). 
}
In order to obtain the minimum value of $Q$,
we adopt the gradient descent algorithm to update all parameters,
\begin{eqnarray}
W_{i} & \leftarrow & W_{i} + l^{W}_{i} \bigg(q_{1} \frac{1}{N} \frac{\partial O(T)}{\partial W_{i}} + q_{2} \frac{\partial \mathcal{L}}{\partial W_{i}} \bigg) , i = 1 , 2 , 3 ,\nonumber \\
B_{i} & \leftarrow & B_{i} + l^{B}_{i} \bigg(q_{1} \frac{1}{N} \frac{\partial O(T)}{\partial B_{i}} + q_{2} \frac{\partial \mathcal{L}}{\partial B_{i}} \bigg) , i = 1 , 2,
\label{eq:variation}
\end{eqnarray}
for the next variation.
Here, the learning rates $l^{W}_{i}$ and $l^{B}_{i}$ are adjustable parameters depending on the impact of the corresponding parameters on $Q$.

The gradients of the $O(T)$ and $\mathcal{L}$ with respect to all the parameters are given by
\begin{eqnarray}
    \frac{\partial O(T)}{\partial W_{i}} &=& \int_{-T}^{T} dt \frac{\delta O(T)}{\delta g(t)} \frac{\partial g(t)}{\partial W_{i}} , \nonumber \\
    \frac{\partial O(T)}{\partial B_{i}} &=& \int_{-T}^{T} dt \frac{\delta O(T)}{\delta g(t)} \frac{\partial g(t)}{\partial B_{i}},
\end{eqnarray}
and
\begin{eqnarray}
    \frac{\partial \mathcal{L}}{\partial W_{i}} =2 (g(-T) - r) \frac{\partial g(t)}{\partial W_{i}} \bigg|_{t = -T} + 2 (g(T) - r) \frac{\partial g(t)}{\partial W_{i}} \bigg|_{t = T}, \nonumber \\
    \frac{\partial \mathcal{L}}{\partial B_{i}}=2 (g(-T) - r) \frac{\partial g(t)}{\partial B_{i}} \bigg|_{t = -T} + 2 (g(T) - r) \frac{\partial g(t)}{\partial B_{i}} \bigg|_{t = T}. \nonumber
\end{eqnarray}
The gradients of $g(t)$ with respect to its parameters are calculated using the chain rule and are then normalized to mitigate the risk of gradient explosion.

We only update the corresponding parameters of the newly increased neurons according to Eq.~(\ref{eq:variation}), while keep the other parameters unchanged.
Next, if both  the observable  $O$ and the loss $\mathcal{L}$, calculated using the updated parameters, are comparatively smaller than those obtained from the original parameters,
we will accept the new neurons as part of the neural network,
namely, $n_{1} \leftarrow n_{1} + {\delta n_{1}}$ and $n_{2} \leftarrow n_{2} + {\delta n_{2}}$.
Otherwise,
the newly introduced units are reinitialized randomly. 
We can iterate through 
the aforementioned steps to obtain the desired control field. 
{Adding more neurons into neural network architectures involves directly with the efficiency and scalability of the model, presenting {numerical cost} as an interesting aspect that warrants further study~\cite{huang2022provably}.
}
{While it is challenging to provide a comprehensive overview due to the complicate interplay between model complexity and computational resources, the current insights suggest that computational time and memory requirements tend to scale nearly linearly with the number of parameters (See more details in appendix~\ref{AppB}).}
It is also important to note that, in order to prevent the neural network from getting trapped in potential local minima, we can establish a criterion for the neural network to  adjust its architecture after an extended period.  Furthermore, we may approve the update even if the observable $O$ or the loss $\mathcal{L}$, calculated using the updated parameters, is slightly higher than those computed with the original parameters.

\section{Optimal suppression of defect generation in the quantum Ising model} \label{optimal}
\subsection{Quantum Ising chain and gradient of the observable}

In this following, we will concretely apply the DCNN method to physical scenarios, 
{specifically focusing on enhancing the suppression of defect generation and improving the preparation of cat-states during the passage across the QCP of the 1D quantum Ising model. For the sake of numerical simplicity, we employ periodic boundary conditions for the spin chain~\cite{PhysRevLett.95.245701, PhysRevLett.109.115703}. While certain local observables, i.e., local longitudinal magnetization, may differ between periodic and open boundary conditions~\cite{Bialonczyk_2020},  it is expected that the global observables, such as defect density and cat-state fidelity,  will demonstrate analogous behavior irrespective of the boundary conditions.}
The Hamiltonian of a periodic quantum Ising chain with $\vec{\sigma}_{N + 1} = \vec{\sigma}_{1}$, is given by
\begin{eqnarray}
    \hat{H}_{\mathrm{QIM}} = - \sum_{j = 1}^{N} [\sigma_{j}^{x} \sigma_{j + 1}^{x} + g(t) \sigma_{j}^{z}],
     \label{eq_QIM}
\end{eqnarray}
where $g(t)$ is a global transverse magnetic field 
that we aim to control, and for simplicity, we assume $N$ to be even.
This model exhibits a quantum phase transition at $g_c = 1$ between the ferromagnetic phase for $0 < g < 1$ and the paramagnetic phase for $g > 1$.
\par The Hamiltonian $\hat{H}_{\mathrm{QIM}}$ can be diagonalized
through a standard procedure, which involves applying the Jordan-Wigner transformation followed by the Fourier transformation on the fermionic operators.
In this study, we are interested in dynamical protocols starting from the paramagnetic phase with $g_i>1$,  ensuring that only the subspace with an even fermion number parity is pertinent~\cite{PhysRevE.101.042108}, yielding
\begin{eqnarray}
\hat{H}^{(+)}_{\mathrm{QIM}} &=& \sum_{k\in K} \hat{H}_k,\nonumber\\
\hat{H}_k&=&2(c^\dag_k,c_{-k}) 
\left(
      \begin{array}{cc}
        g+\cos k & \sin k \\
       \sin k & -g-\cos k \\
     \end{array}
    \right)
\left(
     \begin{array}{c}
       c_k \\
       c^\dag_{-k} \\
     \end{array}
    \right),\nonumber\\
     \label{eq:Hk}
\end{eqnarray}
where $K=\{\pi/N,3\pi/N,\cdots,\pi-\pi/N\}$ and the $c_k$'s are fermionic operators satisfying $\{c_k,c_{k'}\}=0$ and $\{c_k,c^\dag_{k'}\}=\delta_{kk'}$.
{Note that for open boundary conditions, the necessary adjustment to the momentum set $K$ involves solving the corresponding transcendental equation~\cite{PhysRevB.35.7062}.} The ground and excitation states of the mode Hamiltonian $\hat{H}_k$ are given by
\begin{eqnarray}
    | G_{k} \rangle & = & \cos \frac{\theta_{k}}{2} | \mathrm{vac} \rangle_{k} + \sin \frac{\theta_{k}}{2} | k , -k \rangle, \nonumber \\
    | \bar{G}_{k} \rangle & = & \sin \frac{\theta_{k}}{2} |\mathrm{vac}\rangle_{k} - \cos \frac{\theta_{k}}{2} | k , -k \rangle,
\end{eqnarray}
where $| \mathrm{vac} \rangle_{k}$ is the vacuum state of $c_{\pm k}$ and $|k,-k\rangle=c^\dag_k c^\dag_{-k}| \mathrm{vac} \rangle_{k}$ . The angle $\theta_k$ is determined by $\tan \theta_{k} =-\sin k /(g+\cos k)$, 
where $\Lambda_k(t)=2\sqrt{g^2(t)+2g(t)\cos k+1}$ is the single-particle dispersion with which one has $\hat{H}_k|G_k\rangle=-\Lambda_k|G_k\rangle$ and $\hat{H}_k|\bar{G}_k\rangle=\Lambda_k|\bar{G}_k\rangle$. In the present problem, the operator $\hat{Y}=-\sum_j\sigma^z_j$ gives
\begin{eqnarray}
\hat{Y}_k=-2\sum_{k\in K}( c^\dag_k c_k+c^\dag_{-k}c_{-k} -1).
\end{eqnarray}

\par  Defects or domain walls will be inevitably  produced during the transition  from the paramagnetic phase to the ferromagnetic phase. The number of defects is measured by the operator
\begin{eqnarray}
\hat{D}=\frac{1}{2} \sum_{j = 1}^{N} (1 - \sigma_{j}^{x} \sigma_{j + 1}^{x}) = \sum_{k>0} \hat{D}_{k},
\end{eqnarray}
where $\hat{D}_{k}$ is represented by the corresponding matrix form 
\begin{equation}
    \mathcal{D}_{k} = 1 + \left(
        \begin{array}{cc}
             \cos k  &  - \sin k  \\
             - \sin k  & - \cos k
        \end{array}
        \right)
\end{equation}
in the two-dimensional space spanned by $\{|\mathrm{vac}\rangle_k,|k,-k\rangle\}$. 
According to Eq.~(\ref{grad1}), the gradient of $D(T)$ with respect to $g(t)$ is
\begin{eqnarray}
    \frac{\delta D(T)}{\delta g(t)} & = & 4 \textrm{Im} \sum_{k > 0 , k \in K} \langle \phi_{k}(T) | \cos k \sigma_{k}^{z} - \sin k \sigma_{k}^{x} | \bar{\phi}_{k}(T) \rangle \nonumber \\
    & &\langle \bar{\phi}_{k}(t) | \sigma_{z}^{k} | \phi_{k}(t) \rangle.
    \label{eq:dD_dg}
\end{eqnarray}

\subsection{DCNN guided by the final defect density}

{We consider a quench from ${g(-T)} = 2$ to ${g(T)} = 0$ within the time interval $[-T , T]$.
 We focus on a power-law quench characterized by the function  $g(t) = 1 - |t / T|^{r} \textrm{sgn} (t)$, which has been demonstrated to effectively reduce the final defect density, defined as $\rho(T)\equiv D(T)/N$~\cite{PhysRevB.91.041115}. 
 Notably, this power-law quench strategy surpasses the conventional linear quench~\cite{PhysRevLett.95.245701}, the assisted adiabatic passage~\cite{PhysRevLett.109.115703} and the local adiabatic evolution approach~\cite{PhysRevA.65.042308} in minimizing $\rho(T)$.}
To test the DCNN method, we firstly consider the case of $T = 50$ and $N = 50$.
 In the initial neural network, there are two hidden layers with $n_1=10$ neurons in the first layer and $n_2=25$ neurons in the second layer.
To improve the learning efficiency of the neural network, we employ pre-training based on a gradient-oriented power-law quench to establish the initial parameters.
We then set ${\delta n_{1}} = 5$ and ${\delta n_{2}} = 10$. 
These values are adaptable and can be increased throughout the learning process. 
The parameters for the newly added neurons in these layers are randomly initialized, following a uniform distribution ranging from $-0.1$ to $0.1$. 
We set the learning rates to be $l^{W}_{i} = l^{B}_{i} = 0.05$  for $\ i = 1 , 2$ and $l^{W}_{3} = 0.5$.
Due to the significant impact of $W_3$ on the $Q$, the learning rate of $W_{3}$ 
is set at a value tenfold higher than that of the other parameters.
The weights assigned to $Q$ are $q_{1} = 5$ and $q_{2} = 1$, and we ensure the normalization of gradients throughout the learning process. 

\par
\begin{figure}
    \includegraphics[width=\columnwidth]{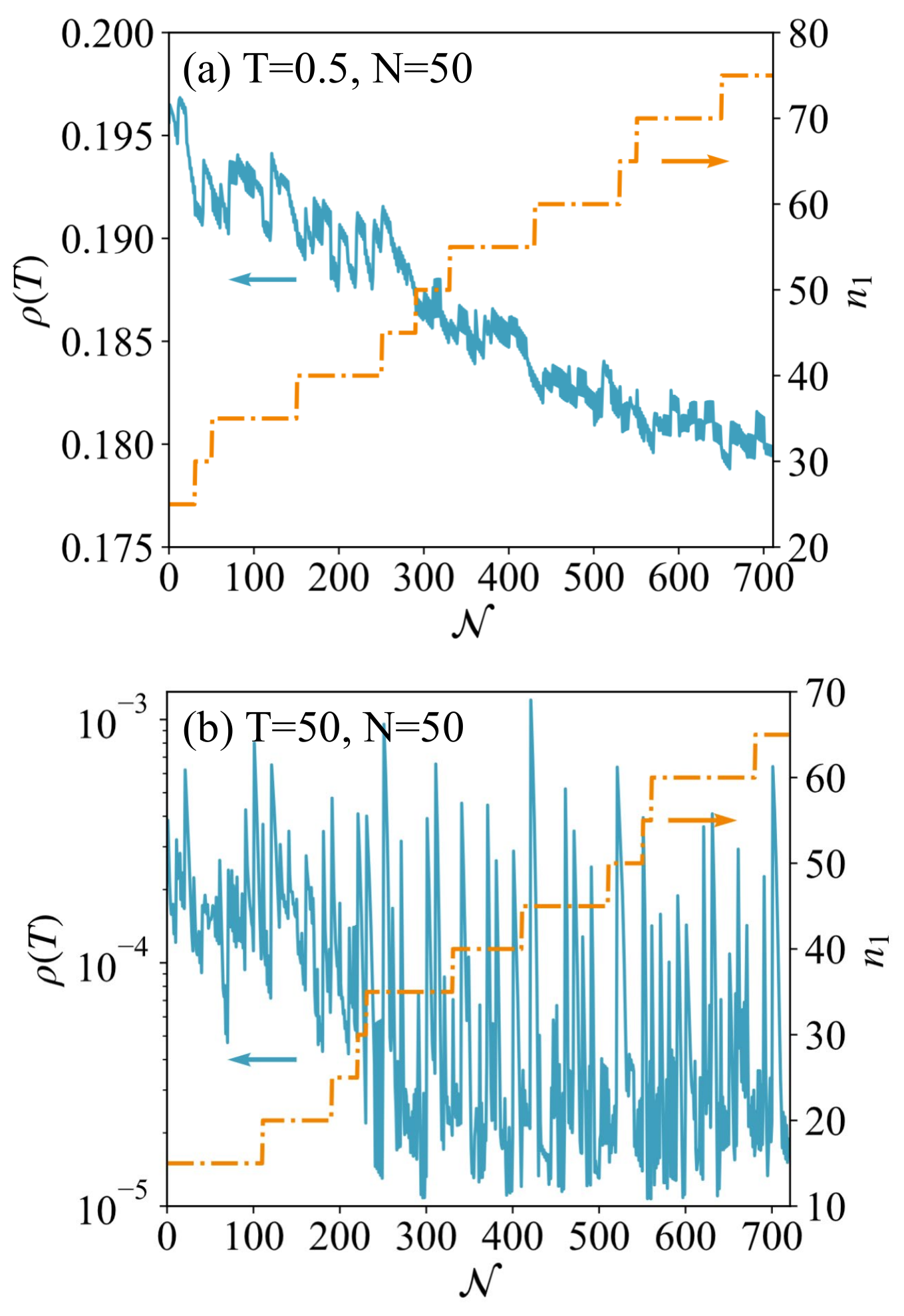}
    \caption{The learning results of the DCNN for (a) $T = 0.5, N = 50$ and (b) $T = 50 , N = 50$.
    The number of neurons in the first hidden layer $n_{1}$ (orange dash-dot line) illustrates the change of the structure of the neural network with the learning iterations $\mathcal{N}$.
    The blue line shows the specific optimization process of the {final} defect density $\rho(T)$.
    }
    \label{fig_learning}
\end{figure}

 The learning results of the DCNN for $T = 0.5 , N = 50$ and $T = 50 , N = 50$ are shown in Fig.~\ref{fig_learning}.
The number of neurons in the first layer $n_{1}$ 
exhibits a stepwise increase during the learning process. 
Each step signifies a transformation in the neural network's scale, leading to an enlargement of the model's representational space. During these steps, new nodes are introduced with different initializations. The primary objective of this process is to incrementally decrease the value of $\rho(T)$ through successive modifications of $g(t)$ facilitated by these additional nodes. 
{The interplay between reducing the final defect density and the system's evolution through two fixed points presents an intriguing scenario. In certain cases, reducing $\rho(T)$ may result in the control field $g(t)$ diverging more substantially from these fixed points. To accelerate the convergence rate of the neural networks, we will selectively abandon the optimized form of the {control} field, although the trade-off could potentially become less pronounced as the neural network's learning process advances.
In this context, the DCNN exhibits a pronounced ability to navigate multiple objectives and thereby effectively mitigates oscillatory behaviors between competing objectives, fostering a more rapid convergence. }
In Fig.~\ref{fig_learning}, we observe an oscillatory decrease in $\rho(T)$ with increasing learning iteration number $\mathcal{N}$, indicating significant modifications to the neural network due to the introduction of additional modes.
Note that outliers resulting from significant modifications in the learning process should not be a concern, as the {optimization strategy} automatically discards these values.
As the learning process progresses in both scenarios, $\rho(T)$ eventually exhibits convergence with minimal fluctuations. Specifically,    $\rho(T)$ converges to a finite value for $T=0.5$, while for $T=50$ it stabilizes around a minimal value.

\subsection{Numerical results and discussions}
\begin{figure}
    \includegraphics[width=\columnwidth]{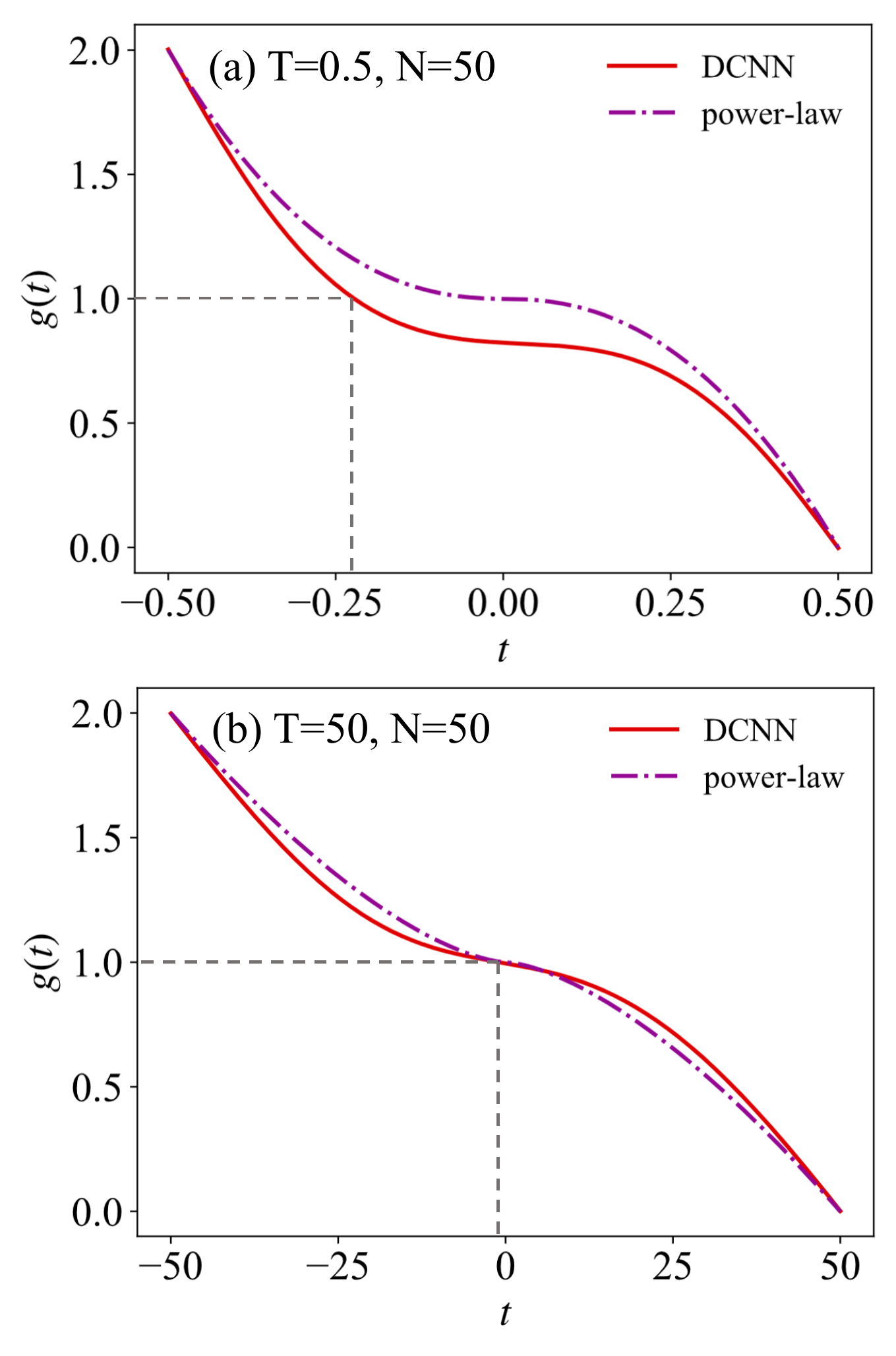}
    \caption{The comparison between the optimized forms of the control field $g(t)$ based on the DCNN (red solid line) and the gradient-based power-law quench (purple dash-dot line) for (a) $T = 0.5, N = 50$ and (b) $T = 50 , N = 50$.
    {In both panels, the horizontal grey dashed lines indicate QCPs, while the grey dashed vertical lines identify $t^*$.  
    }}
    \label{fig_g(t)}
\end{figure}

In Fig.~\ref{fig_g(t)}, we present the optimized control field $g(t)$ obtained by both the DCNN method and the gradient-based power-law quench. 
In contrast to the power-law control, the resulting $g(t)$ lacks symmetric structure and passes through the QCP at a specific time $t^* (< 0)$.  Interestingly, we observe that $t^*$ tends to increase as $T$ increases. For the long-term evolution, it remains imperative to navigate through the QCP as slowly as feasible. 
Nevertheless, in both instances, the slowest rate of change in $g(t)$ occurs at $t = 0$. 
{These characteristics of $g(t)$ could provide insights into the optimal timing and approach for traversing the critical point, thereby minimizing defect generation in practical pulse design applications. Furthermore, such insights may hold the potential for generalization to state preparation protocols across various quantum critical systems.}

\begin{figure}
    \includegraphics[width=\columnwidth]{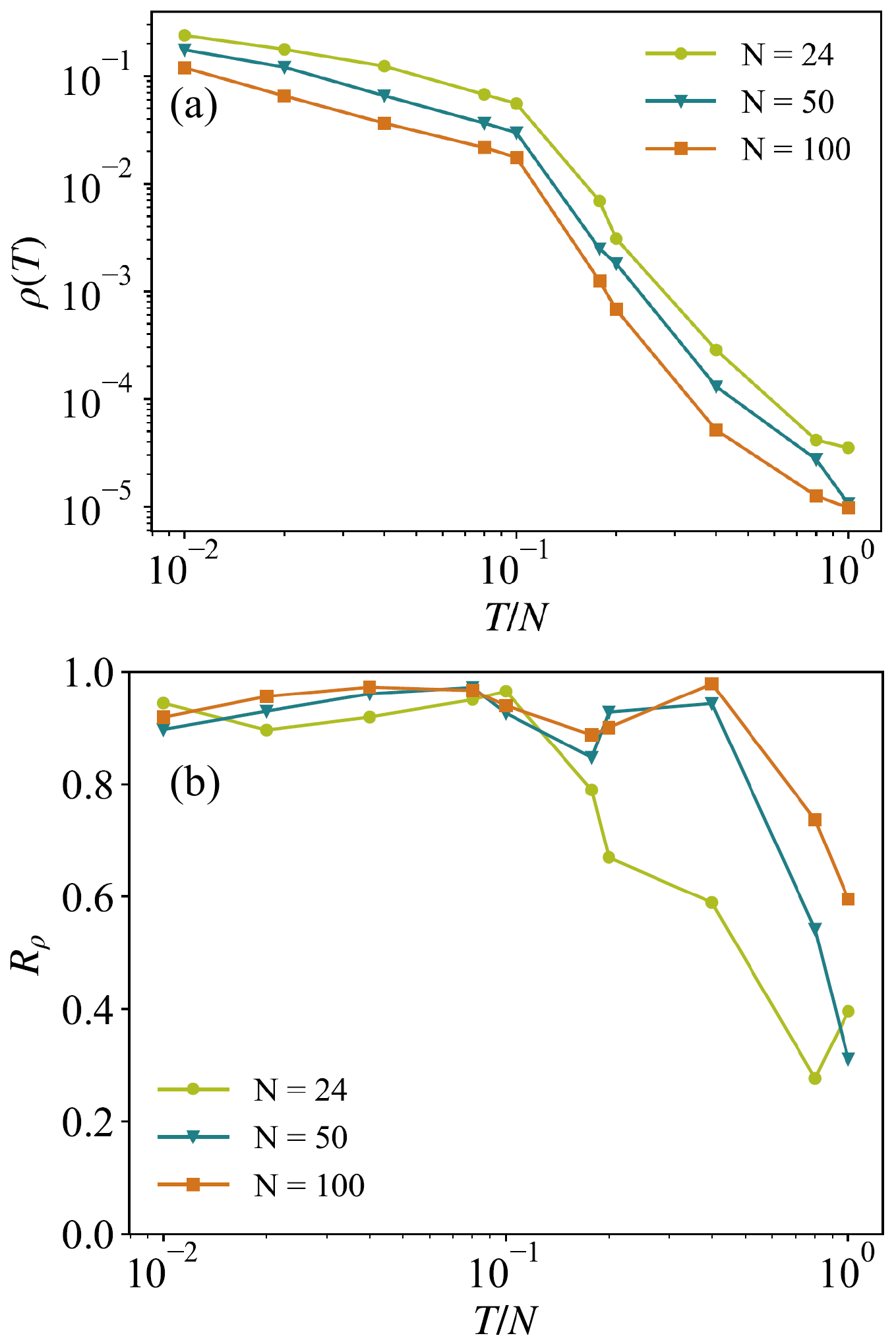}
    \caption{(a) The {final} defect density $\rho(T)$ with respect to different $T$ and $N$ under the optimization by the DCNN.
    (b) The ratio $R_{\rho}$ of the {final} defect density optimized by the DCNN relative to that obtained by the gradient-based power-law quench method.}
    \label{fig_rho}
\end{figure}

The optimized final defect density $\rho(T)$ using the DCNN is plotted in Fig.~\ref{fig_rho}(a) for various system sizes with $T/N$ ranging from 0.01 to 1. 
To assess the relative effectiveness of different optimization methods, we plot in Fig.~\ref{fig_rho}(b) the ratio $R_{\rho}$ of
the defect density optimized by the DCNN to {that} achieved through the gradient-based power-law quench. 
We see that the DCNN outperforms the optimal power-law quench across a broad range of parameters, particularly for small-sized systems and in the long-duration limit.
Similar to the case of the optimal power-law profile, we observe a crossover in the scaled time $T/N$ beyond which the optimized defect density undergoes a sharp decline. The simultaneous occurrence of this crossover for different values of $T$ and $N$ suggests that the crossover time $T_{\rm c}$ is proportional to $N$, contrasting with the Kibble-Zurek scale for the linear quench, where $T_{\rm KZ}$ scales as $N^2$~\cite{PhysRevLett.95.245701}. 
This occurrence can be ascribed to the implications of the quantum speed limit, which establishes a connection between the maximum speed of evolution and the system’s energy uncertainty and mean energy. 
Prior studies suggest that the quantum evolution in a many-body model primarily involves two lowest states in the scenario of small excitations~\cite{PhysRevApplied.14.044043, bera_emergent_2022, PhysRevA.94.012328}. This enables effective mapping into two-level Hamiltonians~\cite{bason_high-fidelity_2012, PhysRevLett.114.233602, li_optimal_2023, aau5999}.
In the even subspace of $k$ mode in the 1D transverse Ising model,
the mode Hamiltonian $\hat{H}_k$ in Eq. (\ref{eq:Hk}) can be reformulated in a Landau-Zener-type form as $\hat{H}_{k} = \gamma_k \sigma_k^{z} + \omega_k \sigma_k^{x}$ with $\gamma_k=2(g + \cos k)$ and $\omega_k = - 2 \cos k$.
There exists an intrinsic quantum speed limit $T_{\rm QSL}$ to drive a general initial state to a final desired state~\cite{PhysRevLett.111.260501}, expressed by
\begin{eqnarray*}
\tan 2 \omega_k T_{\rm QSL} \propto \frac{1}{\omega_k}.
\end{eqnarray*}
For the quench with $t \in [-T , T]$ in the present scenario, the lowest mode $k_N$ specified by $\omega_{k_N} \approx - 2 \pi / N$ yields $2 \omega_{k_N} T_{\rm QSL} \approx - \pi / 2$ or $T_{\rm QSL} / N \approx 1 / 8$ for sufficiently large $N$, serving as a lower bound for $T_{\rm QSL}$ for the entire system, considering all modes $k$.

\begin{figure}
    \includegraphics[width=\columnwidth]{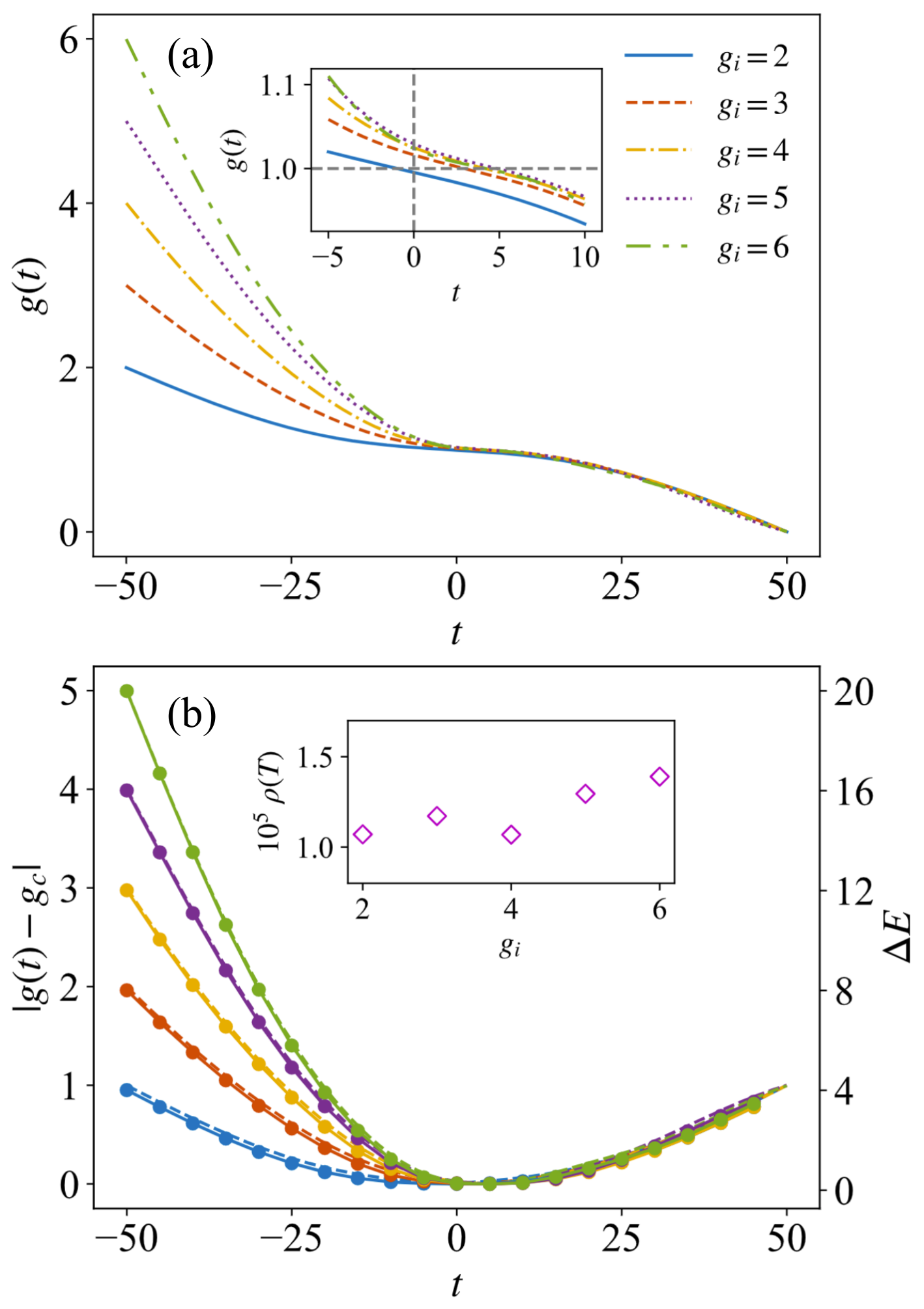}
    \caption{(a)The optimized control fields $g(t)$ obtained by the DCNN protocol with different initial values $g_{i}$ for $T = 50, N = 50$. {Inset magnifies the control fields around $t^*$.
    (b) The absolute value of the relative offset of $g(t)$ to $g_c$ (dashed lines) and the energy gap $\Delta E$ (solid lines with markers) with respect to $t$.
    Inset shows the corresponding final defect density $\rho(T)$ for different $g_{i}$. }}
    \label{fig_gi}
\end{figure}

{The gradient-based power-law quench method, which depends on a specific power-law scheme to pass through fixed points, faces limitations in adaptability across varied initial and final state scenarios. This limitation is particularly pronounced when these states do not exhibit symmetry relative to the critical point, thereby complicating the determination of a suitable power-law function. In contrast, the DCNN offers a more adaptable framework, enabling more straightforward modifications to its optimization strategy. This adaptability suggests that the DCNN approach may provide more nuanced and flexible solutions compared to the gradient-based power-law quench method, especially regarding its application across diverse quenching scenarios.}
We plot the {optimized} control fields with different initial values of $g_i$ in Fig.~\ref{fig_gi} {(a)} for $T = 50$ and $N = 50$. 
It is evident that $g(t)$ smoothly passes through the QCP at some $t^{*}$ around $t=0$ at a {relatively} slow pace. 
As $g_i$ increases while keeping $g_f$ constant, mitigating non-adiabatic effects within a finite duration becomes challenging. 
The {optimized} control fields strikes a balance between  
a deliberate traversal of the QCP 
and avoiding excessive speed in other regions~\cite{PhysRevLett.101.076801}. 
{
To investigate the role of the lowest mode in the optimization process, we plot the absolute value of the relative offset of $g(t)$ to $g_c$ and the energy gap $\Delta E$ in Fig.~\ref{fig_gi} (b). One finds a discernible correlation between the deviation of $g(t)$ from $g_c$ and the energy gap associated with the lowest mode, encapsulated by the approximate relation $|g(t) - g_c| \approx 0.25 \Delta E$.} 

\begin{figure}
    \includegraphics[width=\columnwidth]{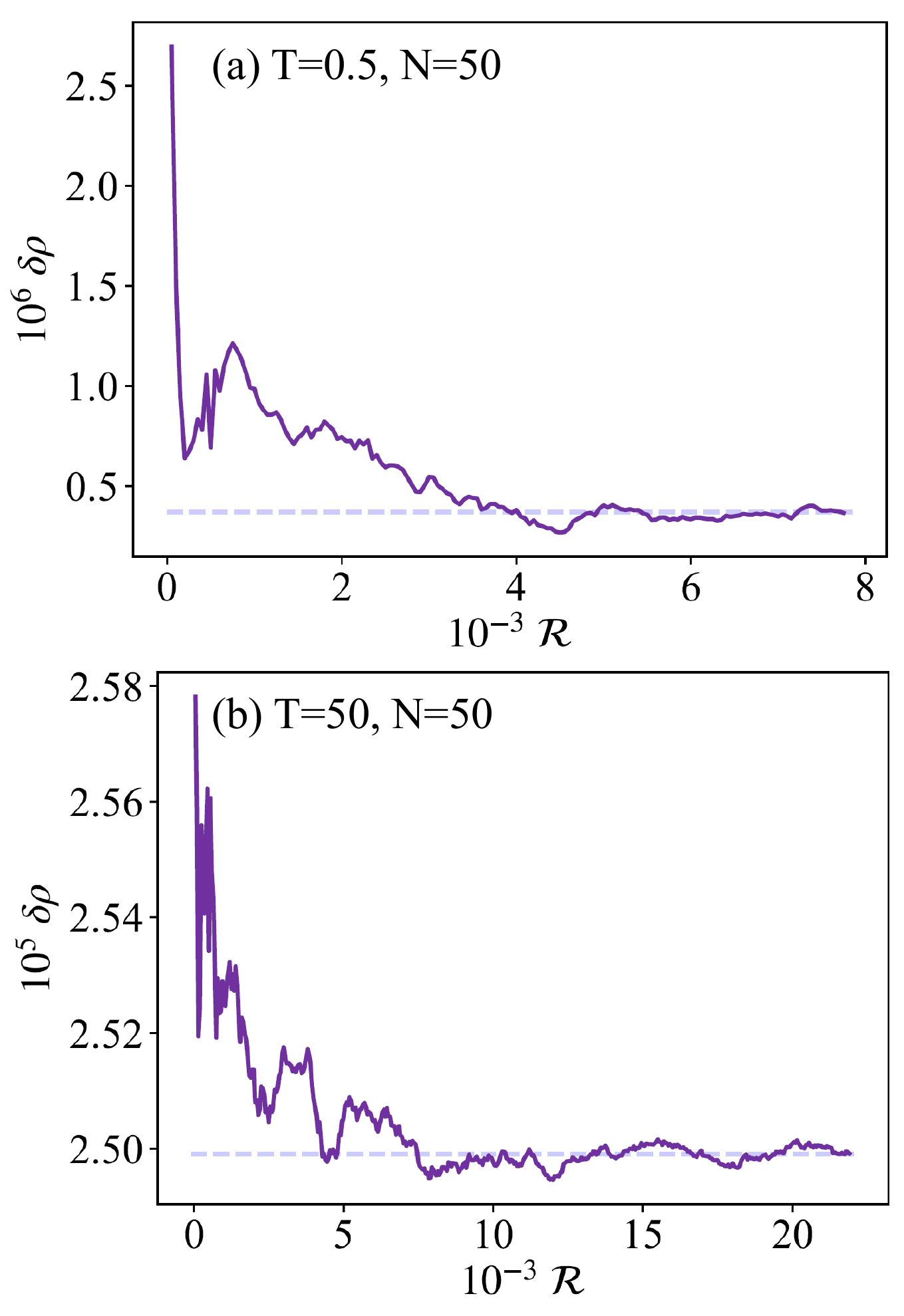}
    \caption{
    {
   Mean relative offsets $\delta \rho$ of the final defect density from the ideal value subjected to white Gaussian noise, as a function of the total number of numerical simulations $\mathcal{R}$, with $\textrm{SNR} = 10$. (a)   $T = 0.5$, $N = 50$, and (b) $T = 50$, $N = 50$. Here, we define $\mathcal{R} = 50p$, where $p = 155$ for (a) and $p = 438$ for (b). Dashed lines in both panels indicate the convergent values.
   } }
    \label{fig_random}
\end{figure}

To explore potential applications, we assess the robustness of our protocol to spin number fluctuations~\cite{PhysRevLett.106.190501}, considering cases with $T = 50, N = 50$ and $T = 0.5, N = 50$ as examples. 
{In experimental implementations, it is inevitable to encounter discrepancies in the actual number of spins present in the system.}
We {thus} introduce a number fluctuation parameter, $\delta N = \pm 4$. The {final} defect density remains below $9.13 \times 10^{-5}$ for $T = 50, N = 50$ and below $0.1798$ for $T = 0.5, N = 50$, indicating the robustness of our proposal against spin number fluctuations.
To delve deeper into the effects of random noise, we examine the additive white Gaussian noise model, denoted as $\mathcal{A}_{G}[g(t), \textrm{SNR}]$. We incorporate it into the control field using the relation:
\begin{equation}
    g_{q}(t) = g(t) + \mathcal{A}_{G}[g(t) , \textrm{SNR}],
\end{equation}
where $q$ represents each random generator $\mathcal{A}_{G}[g(t) , \textrm{SNR}]$, with $\textrm{SNR} = 10 \log_{10} (P_{s} / P_{n})$,
where $P_{s}$ and $P_{n}$ are the power of signal and noise, respectively.
{We evaluate our protocol's robustness against white Gaussian noise by analyzing the mean relative offsets, $\delta \rho$, from the ideal defect densities, conducting $\mathcal{R}$ numerical simulations with $\textrm{SNR} = 10$ for $T = 0.5, N = 50$ and $T = 50, N = 50$. Fig.~\ref{fig_random} depicts $\delta \rho$ versus $\mathcal{R} = 50p$, where $p$ is the simulation index, showing convergence of defect densities to $3.57 \times 10^{-5}$ and 0.1798 for the respective cases. These results highlight our scheme exhibits a high degree of robustness against random noise.}

\section{Optimal cat-state preparation in the quantum Ising model} 
 \label{sec_catstate}
 
\par In this section, we employ the DCNN to tackle a more complex task: optimally preparing a cat state in the quantum Ising chain.  The Schr\"{o}dinger cat states have attracted much attention from researchers due to their fundamental implications.  
The quantum superpositions of two macroscopically distinct states are not only interesting for testing the fundamentals of quantum mechanics such as the quantum-to-classical transition~\cite{schrodinger_gegenwartige_1935, RevModPhys.85.1103}, but
also a valuable resource  for quantum information processing~\cite{zurek_sub-planck_2001, ofek_extending_2016, PhysRevLett.127.087203}.  Noteworthy examples of maximally entangled cat states include  the NOON state~\cite{dowling_quantum_2008} and Greenberger-Horne-Zeilinger (GHZ) state~\cite{PhysRevA.105.062456}, which can push the estimation precision to the Heisenberg limit~\cite{PhysRevLett.96.010401, PhysRevLett.97.150402, PhysRevLett.102.100401}. 
Of particular interest are spin cat states  for use in atomic clocks~\cite{martin_quantum_2013} and magnetometers~\cite{PhysRevLett.104.133601} due to the potential applications in diverse fields from materials science to  medical diagnostics~\cite{le_sage_optical_2013}.  
However, the realization of cat states in practical experiments faces challenges due to their sensitivity to decoherence and the delicate entanglement properties~\cite{leibfried_creation_2005}.  

The target state we aim to achieve is  
\begin{eqnarray}
|\mathrm{cat}\rangle=\frac{1}{\sqrt{2}}(|\rightarrow\cdots\rightarrow\rangle+|\leftarrow\cdots\leftarrow\rangle),
\end{eqnarray}
which is an equally weighted superposition of the two degenerate fully polarized states $|\rightarrow\cdots\rightarrow\rangle$ and $|\leftarrow\cdots\leftarrow\rangle$. Our goal is to identify an {optimized} function $g(t)$ that maximizes the final fidelity, defined as  
\begin{eqnarray}
F(T)=|\langle\phi(T)|\mathrm{cat}\rangle|^2.
\end{eqnarray}
It is important to note that maximizing this fidelity poses a more stringent requirement than simply minimizing the final defect density. This is because any linear combination of $|\rightarrow\cdots\rightarrow\rangle$ and $|\leftarrow\cdots\leftarrow\rangle$  would result in vanishing defects. 

\par It is shown in Ref.~\cite{PhysRevE.101.042108} that $|\mathrm{cat}\rangle$ can be written in the momentum space as a product state
\begin{eqnarray}
|\mathrm{cat}\rangle=\prod_{k\in K}|\chi_k\rangle
\end{eqnarray}
with $|\chi_k\rangle\equiv \sin\frac{k}{2}|\mathrm{vac}\rangle_k+\cos\frac{k}{2}|k,-k\rangle$. Therefore, $F(T)$ is actually the expectation value of the operator $\hat{F}\equiv\prod_{k\in K}|\chi_k\rangle\langle \chi_k|$ in the final state $|\phi(T)\rangle$. From Eq.~(\ref{grad2}), the dependence of $F(T)$ with respect to $g(t)$ is delineated as
\begin{eqnarray}
    \frac{\delta F(T)}{\delta g(t)} & = & 2 \textrm{Im} \prod_{k} |\langle \phi_{k}(T) | \chi_{k} \rangle|^{2} \nonumber \\
    & & \sum_{k'} \frac{\langle \chi_{k} | \bar{\phi}_{k'}(T) \rangle \langle \bar{\phi}_{k'}(t) | \hat{Y}_{k'} | \phi_{k'}(t) \rangle}{\langle \chi_{k} | \phi_{k'}(T) \rangle}.
\end{eqnarray}

\par
\begin{figure}
    \includegraphics[width=\columnwidth]{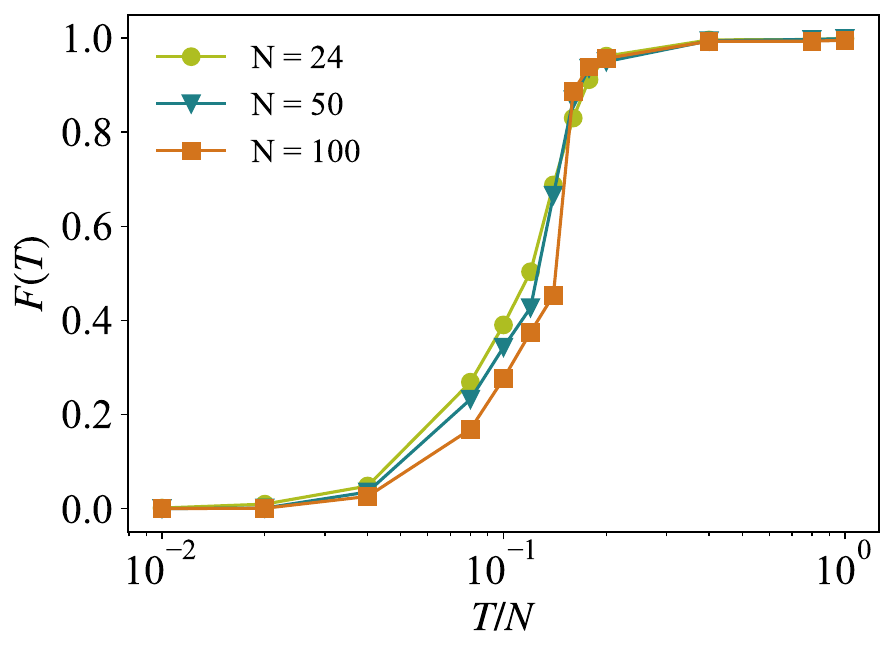}
    \caption{The fidelity {$F(T)$} of the cat state is evaluated under various values of  $T$ and $N$, optimized by the DCNN.
    }
    \label{fig_f}
\end{figure}

 Within the DCNN framework, we also examine the quench from $g(-T)= 2$ to $g(T) = 0$ over the time interval $[-T , T]$. The objective function is $Q = q_{1} F(T) + q_{2} \mathcal{L}$. 
The update of all parameters given by Eq.~(\ref{eq:variation}) will undergo corresponding adjustments to reflect changes in $Q$. 
The optimized fidelity results $F(T)$ {obtained through the DCNN} for sizes $N=24,50,100$ and the ratio $T / N$ ranging from 0.01 to 1 are presented in Fig.~\ref{fig_f}.  
As expected, the achieved fidelity increases as the duration $T$ increases and approaches unity for larger $T/N$. In the intermediate quench time range, specifically between  $0.14 < T/N < 0.18$, a notable surge in $F(T)$ is observed, which is accompanied by a marked decrease in {final} defect density, as illustrated in Fig.~\ref{fig_rho}.   It is noteworthy that this rise in $F(T)$  becomes more pronounced with the increase in system size.
For shorter durations, the fidelity always remains extremely low, a consequence of the inherent constraints imposed by the quantum speed limit.

\begin{figure}
    \includegraphics[width=\columnwidth]{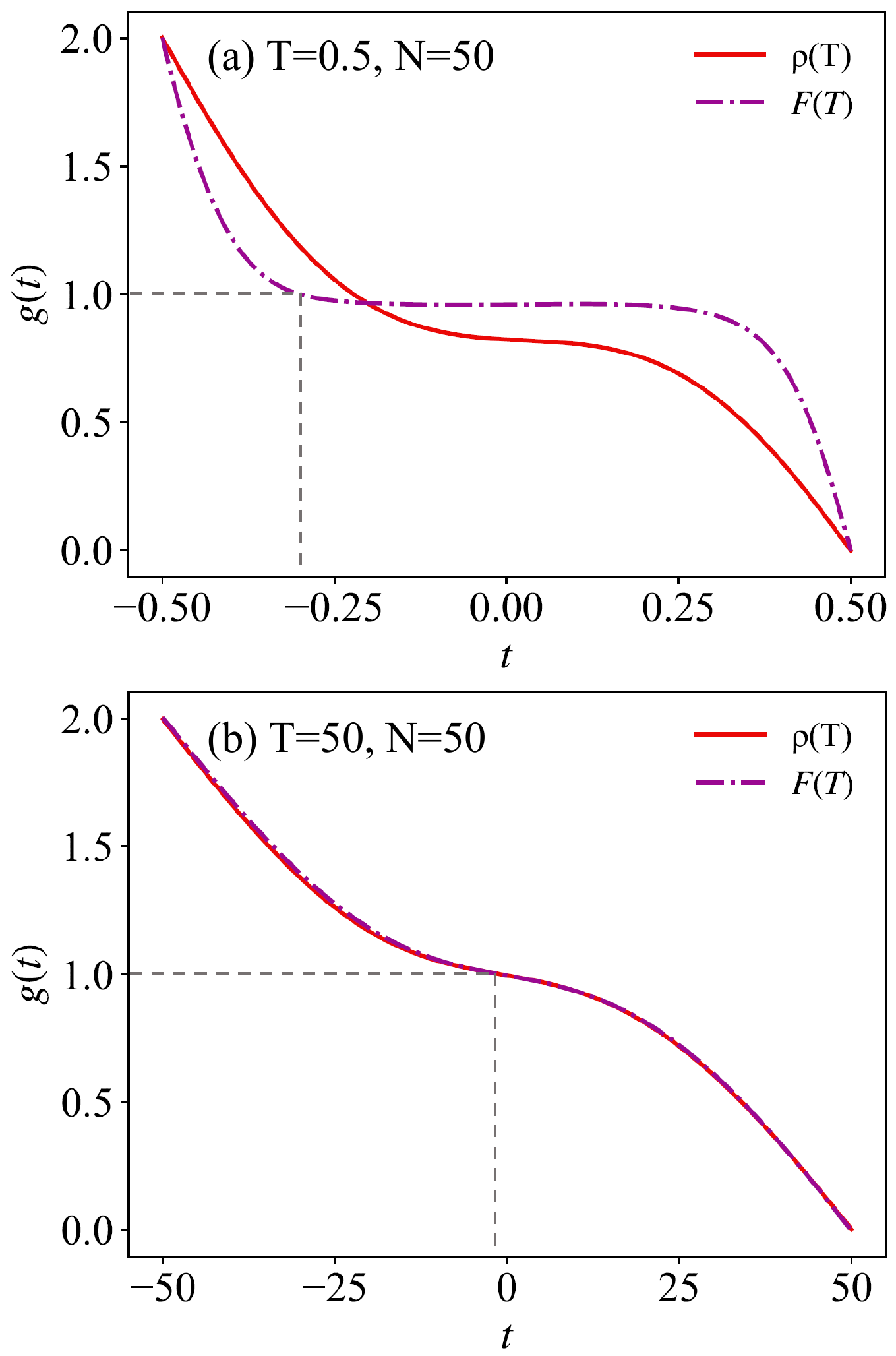}
    \caption{Comparison of the control fields $g(t)$ optimized for the cat-state fidelity $F(T)$ (purple dash-dot line) and the {final} defect density $\rho(T)$ (red solid lines) with (a) $T = 0.5 , N = 50$ and (b) $T = 50, N = 50$. 
    {In both panels, the horizontal grey dashed lines indicate QCPs, while the grey dashed vertical lines identify $t^*$ of $g(t)$ guided by the cat-state fidelity. }
    }
    \label{fig_compare}
\end{figure}

Fig.~\ref{fig_compare} displays the {optimized} control fields $g(t)$, obtained either by maximizing cat-state fidelity $F(T)$ or by minimizing final defect density $\rho(T)$. In Fig.~\ref{fig_compare}(a), a notable difference is observed between the two control fields for short durations. Specifically, the $g(t)$ that minimizes {final} defect density leads to considerably lower fidelity. Conversely, for longer durations, the two profiles are nearly identical in  Fig.~\ref{fig_compare}(b), with each leading to either very low {final} defect density or significantly high fidelity.
This observation {implies} a complex relationship between {final} defect density and state fidelity, indicating that they are not straightforwardly correlated.

\section{Conclusion and outlook} \label{conclusion}

In this study, we introduce a dynamic control neural network (DCNN) approach to suppress the {final} defect density during a passage across a quantum critical point (QCP). Utilizing the dynamic architecture of the neural network combined with a strategic control mechanism, 
we successfully generate control fields that navigate through two fixed points while optimizing the desired objective. By considering various evolution time $T$ and system sizes $N$, our method achieves a lower {final} defect density compared to traditional gradient-based power-law quench methods.
Meanwhile, our study reveals that the optimized {final} defect density undergoes a crossover at a specific critical ratio of quench duration to system size. This transition closely aligns with the quantum speed limit for evolution, which is profoundly influenced by the lowest mode.  We numerically demonstrate that our protocol is robust against the random noise and the spin number fluctuations. 
{Additionally, we demonstrate the effectiveness of our method using control fields that are optimized across a range of initial values.}
By analyzing variational control field forms and comparing them to power-law quench methods, we offer insights for experimental pulse design. 
The versatility of the DCNN is further highlighted by its application to enhancing cat-state fidelity, providing similar optimized solutions indicative of the quantum speed limit. Our findings reveal a non-one-to-one correspondence between {final} defect density and fidelity. To attain high-quality quantum many-body states, it is crucial to exceed the critical ratio of quench duration to system size.

Noting that the DCNN method introduced in this study extends beyond the preparation of spin cat states,
holding promise for a broad range of applications in quantum technologies. 
With the rapid advancements in quantum computation
~\cite{PhysRevLett.131.170601, PhysRevLett.130.106701, PRXQuantum.4.020303, PRXQuantum.4.030323}
and quantum metrology~\cite{PhysRevLett.130.240803, PRXQuantum.4.040305, PRXQuantum.4.040336, PhysRevLett.130.070803}, the DCNN's capability for optimizing complex quantum systems positions it as a valuable tool for enhancing precision in quantum information science.
{
The application of the DCNN method to multi-objective optimization in quantum control demonstrates its potential applicability, especially in complex scenarios such as quantum many-body systems and open quantum systems.}

\begin{acknowledgments}
This work is supported by the National Natural Science Foundation of China (NSFC) under Grant No. 12174194  and stable supports for basic institute research under Grant No. 190101.  N. W. was supported by the National Key Research and Development Program
of China under Grant No. 2021YFA1400803.
\end{acknowledgments}

\appendix

\section{The gradients of $O(T)$ with respect to $g(t)$}\label{AppA}

The gradients of the final expectation value $O(T)$ with respect to the {control} field $g(t)$ can be written as 
\begin{eqnarray}\label{dOdg}
\frac{\delta O(T)}{\delta g(t)}=2\Re\langle G(g_i)|U^\dag(T,-T)\hat{O}\frac{\delta U(T,-T)}{\delta g(t)}|G(g_i)\rangle, \quad
\end{eqnarray}
where
\begin{eqnarray}\label{dUdg}
\frac{\delta U(T,-T)}{\delta g(t)}&=&-iU(T,-T)U^\dag(t,-T)\frac{\partial H(t)}{\partial g(t)}U(t,-T)\nonumber\\
&=&-iU(T,-T)U^\dag(t,-T)\hat{Y}U(t,-T).
\end{eqnarray}
Combining Eq.~(\ref{dOdg}) with Eq.~(\ref{dUdg}), we derive the following expression:
\begin{eqnarray}\label{dOdg1}
\frac{\delta O(T)}{\delta g(t)}=2\Im\langle G(g_i)|\hat{O}(T)\hat{Y}(t)|G(g_i)\rangle.
\end{eqnarray}
For simplicity, the $|G(g_i)\rangle$ is abbreviated as $| G_{i} \rangle$.
We have assumed that the control term $\hat{Y}$ can be expressed as a summation over even operators of independent mode $\hat{Y} = \sum_{\vec{k}} \hat{Y}_{\vec{k}}$. If $O(T)$  can be represented as a sum of even operators over independent modes, such that $\hat{O}(T) = \sum_{\vec{k}} \hat{O}_{\vec{k}}(T)$, then it follows that: 
\begin{eqnarray}
    & &\langle G_{i} | \hat{O}(T) \hat{Y}(t) | G_{i} \rangle \nonumber\\
    & = & \sum_{\vec{k} , \vec{k}'} \langle G_{i , \vec{k}} | \hat{O}_{\vec{k}}(T) \hat{Y}_{\vec{k}'}(t) | G_{i , \vec{k}'} \rangle \nonumber \\
    & = & \sum_{\vec{k}' \not= \vec{k}} \langle G_{i , \vec{k}} | \hat{O}_{\vec{k}}(T) \hat{Y}_{\vec{k}'}(t) | G_{i , \vec{k}'} \rangle + \sum_{\vec{k}} \langle G_{i , \vec{k}} | \hat{O}_{\vec{k}}(T) \hat{Y}_{\vec{k}}(t) | G_{i , \vec{k}} \rangle \nonumber \\
    & = & \sum_{\vec{k} , \vec{k}'} \langle G_{i , \vec{k}} | \hat{O}_{\vec{k}}(T) | G_{i , \vec{k}} \rangle \langle G_{i , \vec{k}'} | \hat{Y}_{\vec{k}'}(t) | G_{i , \vec{k}'} \rangle \nonumber \\
     & - & \sum_{\vec{k}} \langle G_{i , \vec{k}} | \hat{O}_{\vec{k}}(T) | G_{i , \vec{k}} \rangle \langle G_{i , \vec{k}} | \hat{Y}_{\vec{k}}(t) | G_{i , \vec{k}} \rangle  \nonumber \\
     & + & \sum_{\vec{k}} \langle G_{i , \vec{k}} | \hat{O}_{\vec{k}}(T) \hat{Y}_{\vec{k}}(t) | G_{i , \vec{k}} \rangle \nonumber \\
    & = & \sum_{\vec{k}} \langle G_{i , \vec{k}} | \hat{O}_{\vec{k}}(T) | \bar{G}_{i , \vec{k}} \rangle \langle \bar{G}_{i , \vec{k}} | \hat{Y}_{\vec{k}'}(t) | G_{i , \vec{k}} \rangle \nonumber \\
    & = & \sum_{\vec{k}}\langle\phi_{\vec{k}}(T)|\hat{O}_{1,\vec{k}}|\bar{\phi}_{\vec{k}}(T)\rangle\langle\bar{\phi}_{\vec{k}}(t)|\hat{Y}_{\vec{k}}|\phi_{\vec{k}}(t)\rangle.
    \label{eq:partial_O}
\end{eqnarray}
In deriving the penultimate line of Eq.~\eqref{eq:partial_O},
we have used the identity 
\begin{eqnarray}
\mathbb{I}_{\vec{k}} = | G_{i , \vec{k}} \rangle \langle G_{i , \vec{k}} | + | \bar{G}_{i , \vec{k}} \rangle \langle \bar{G}_{i , \vec{k}} |.
   \label{eq:identity}
\end{eqnarray}
Then, we can obtain Eq.~\eqref{grad1}.
If $O(T)$ can be expressed as a product over even operators of independent modes $\hat{O}(T) = \prod_{\vec{k}} \hat{O}_{\vec{k}}(T)$,
it can be calculated as
\begin{eqnarray}
   && \frac{\delta O_{2}(T)}{\delta g(t)} \nonumber \\ &=&
   2 \Im \sum_{\vec{k}'} \langle G_{i , \vec{k}'} | \hat{O}_{2 , \vec{k}'}(T) \hat{Y}_{\vec{k}'}(t) | G_{i , \vec{k}'} \rangle  \prod_{\vec{k} \not= \vec{k}'} \langle G_{i , \vec{k}} | \hat{O}_{2 , \vec{k}}(T) | G_{i , \vec{k}} \rangle \nonumber \\
    & = & 2 \Im \prod_{\vec{k}} \langle G_{i , \vec{k}} | \hat{O}_{2 , \vec{k}}(T) | G_{i , \vec{k}} \rangle \sum_{\vec{k}'} \frac{\langle G_{i , \vec{k}'} | \hat{O}_{2 , \vec{k}'}(T) \hat{Y}_{\vec{k}'}(t) | G_{i , \vec{k}'} \rangle}{\langle G_{i , \vec{k}'} | \hat{O}_{2 , \vec{k}'}(T) | G_{i , \vec{k}'} \rangle} \nonumber\\
    & = & 2 \Im \prod_{\vec{k}} \langle G_{i , \vec{k}} | \hat{O}_{2 , \vec{k}}(T) | G_{i , \vec{k}} \rangle \nonumber \\ &\times& \sum_{\vec{k}'} \frac{\langle G_{i , \vec{k}'} | \hat{O}_{2 , \vec{k}'}(T) | \bar{G}_{i , \vec{k}'} \rangle \langle \bar{G}_{i , \vec{k}'} | \hat{Y}_{\vec{k}'}(t) | G_{i , \vec{k}'} \rangle}{\langle G_{i , \vec{k}'} | \hat{O}_{2 , \vec{k}'}(T) | G_{i , \vec{k}'} \rangle} \nonumber \\
    & = & 2 \Im \prod_{\vec{k}} \langle \phi_{\vec{k}}(T) | \hat{O}_{2 , \vec{k}} | \phi_{\vec{k}}(T) \rangle \nonumber \\ &\times& \sum_{\vec{k}'} \frac{\langle \phi_{\vec{k}'}(T) | \hat{O}_{2 , \vec{k}'} | \bar{\phi}_{\vec{k}'}(T) \rangle \langle \bar{\phi}_{\vec{k}'}(t) | \hat{Y}_{\vec{k}'} | \phi_{\vec{k}'}(t) \rangle}{\langle \phi_{\vec{k}'}(T) | \hat{O}_{2 , \vec{k}'} | \phi_{\vec{k}'}(T) \rangle}.
\end{eqnarray}
 We also use the identity in Eq.(\ref{eq:identity}) in the above derivation. 

\section{The numerical cost in neural network expansion} ~\label{AppB}
\begin{figure}[ht]
    \includegraphics[width=\columnwidth]{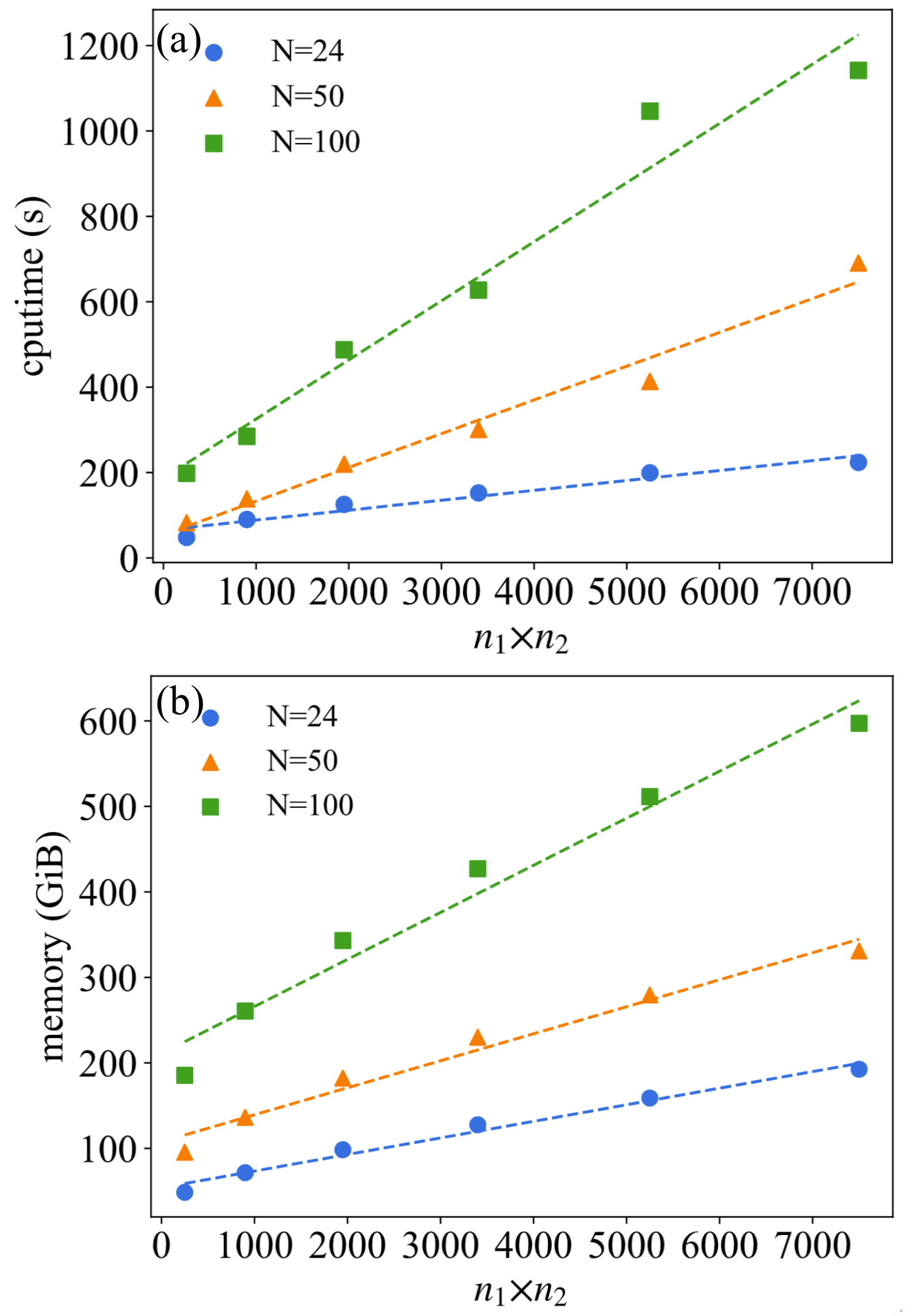}
    \caption{
    (a) Computational time and (b) memory requirement for varying neural network scales with spin lengths of $N = 24$, $N=50$, and $N=100$. 
     The notation $n_1 \times n_2$ denotes a network configuration with $n_1$ and $n_2$ neurons in the first and second hidden layers, respectively. Each point represents the average computational time or memory usage derived from ten iterations of neuron additions to networks of identical scale. The dashed lines indicate linear fits to the data points.}
    \label{fig_cost}
\end{figure}
The parameter scale of expanding a DCNN can be quantified as  $O((n_1 + M \delta n_1)(n_2 + M \delta n_2))$, where $M$ represents the number of expanding the neural network as the network approaches convergence. 
For an intuitive understanding, when the initial network size is substantially  smaller than $M$, the parameter growth simplifies to  $O(M^2)$. Conversely, for an initially larger network relative to $M$, the complexity remains closer to $O(n_1 n_2)$.
For simplicity, we only consider the computational time and the memory consumption of increasing the neurons one time,  transitioning the network from an $n_1 \times n_2$ to an $(n_1 + \delta n_1) \times (n_2 + \delta n_2)$ structure. 
where $n_1 \times n_2$ denotes a network with $n_1$ neurons in the first  and $n_2$ in the second hidden layers, respectively. The overall impact on computational time and memory can be approximated by $\sum_{i} x_i y_i$, with $x_i$ representing the number of iterations until successful neuron integration, and $y_i$ capturing the associated time or memory costs. It's noteworthy that computational duration is subject to a variety of external factors, including CPU performance and specific elements of the matrix.
Assuming a constant spin length, the variation in numerical costs linked to different network sizes is illustrated in Fig.~\ref{fig_cost}. With the parameter scale increases, we observe an almost linear surge in computational time, indicative of the added complexity and data processing requirements. Memory usage similarly rises, though its growth trajectory is shallower compared to computational time,  which suggests a higher data throughput and computational complexity for larger networks.  An extension in spin length further accentuates these resource demands. However, a quantifiable relationship between network size expansion and computational resource consumption requires further investigation.

\normalem
\bibliographystyle{apsrev4-2}
\bibliography{prorefs}

\end{document}